\documentclass[preprint]{aa}
\usepackage{txfonts}
\usepackage{epsfig}
\usepackage{graphicx}

\begin{document}

\title{Chandra Observation  of   the Dipping
  Source XB 1254--690 }
  \subtitle{}

  \author{R. Iaria\inst{1}, T. Di Salvo\inst{1}, G. Lavagetto\inst{1},
    A. D' A{\'\i}\inst{1}, N. R. Robba\inst{1}}
   \offprints{R. Iaria, iaria@fisica.unipa.it}
   \institute{Dipartimento   di  Scienze   Fisiche   ed  Astronomiche,
              Universit\`a  di  Palermo,  via  Archirafi  36  -  90123
              Palermo,  Italy} 
 \date{Received / Accepted}

\authorrunning{R. Iaria  et al.} 
\titlerunning  {Chandra Observation  of   the Dipping
  Source XB 1254--690}

\abstract{ We present the results  of a 53 ks long Chandra observation
  of the dipping source  XB 1254--690.  During the observation neither
  bursts or dips were observed.   From the zero-order image we estimated
  the precise X-ray coordinates of  the source with a 90\% uncertainty
  of 0.6\arcsec.   Since the lightcurve did not  show any significant
  variability, we  extracted the  spectrum corresponding to  the whole
  observation.  We  confirmed the
  presence  of the  \ion{Fe}{xxvi}  K$_\alpha$ absorption  lines with  a
  larger  accuracy   with  respect  to   the  previous  XMM   EPIC  pn
  observation.  Assuming that the line width were due to a bulk motion
  or a turbulence associated to the coronal activity, we estimate that
  the  lines were  produced  in a  photoionized  absorber between  the
  coronal  radius   and  the  outer   edge  of  the   accretion  disk.
}
\maketitle

\section{Introduction}
About 10  Low Mass  X-ray Binaries (LMXB)  are known to  show periodic
dips in  their X-ray  light curves  and most of  them also  show X-ray
burst activity.  The dip intensities,  lengths and shapes  change from
source  to source,  and, for  the same  source, from  cycle  to cycle.
Among the dipping sources, XB 1254­-690 (4U 1254--69), that also shows
type-I X-ray bursts, is peculiar  because it showed a cessation of its
regular  dipping activity (Smale  \& Wachter  1999).  This  source was
observed for  the first  time by  HEAO 1 A2  instrument (Mason  et al.
1980)  when an optical  burst of  $\sim 20$  s duration  occurred. The
optical  counterpart  to  XB  1254-­690,  GR Mus,  was  identified  by
Griffiths et al. (1978) to be a faint blue object (V=19.1), implying a
ratio of X-ray to optical  luminosity $L_x /L_{opt} \sim 200$, similar
to those of other compact X-ray binaries with Population II companions
(Lewin \& Joss, 1983).

EXOSAT observations showed the presence of type-I X-ray bursts and led
to the  discovery of recurrent X-ray  dips with a period  of $3.88 \pm
0.15$  hr (Courvoisier,  Parmar \&  Peacock 1984;  Courvoisier  et al.
1986).  The duration of the dips was 0.8 hrs, representing 20\% of the
binary  cycle, with a  reduction of  the X-ray  flux between  20\% and
90\%.   During the  dips,  X-ray variabilities  were  present on  time
scales  1­-300 s.   This  complex time  structure  indicates that  the
obscuring region  was clumpy and  of variable size. A  modulation with
the same period  of the X-ray dips was also  observed in optical light
curves, with the optical minimum  occurring 0.2 cycles after the X-ray
dip (Motch et al. 1987).  The presence of dips and the lack of eclipse
of the X-ray source gave a  constraint on the inclination angle of the
source that  is between 65$^{\circ}$ and  73$^{\circ}$ (Courvoisier et
al.  1986, Motch et al.  1987).  X-ray bursts and X-ray dips were also
present in  a Ginga observation performed  on August 1990  (Uno et al.
1997).   On  the other  hand  the  dips were  not  present  in a  RXTE
observation  in 1997 between  April 28  and May  1 (Smale  \& Wachter,
1999).   The cessation  of  the dipping  activity  indicated that  the
angular size of the disk edge  (at the impact point with the accretion
stream)   decreased   from   $17^{\circ}-25^{\circ}$  to   less   than
10$^{\circ}$,  moreover  there  was  no  evidence  of  a  simultaneous
variation of  the accretion rate or  the location of the  outer rim of
the  accretion disk  (Smale \&  Wachter,  1999).  The  dips were  also
absent during  a BeppoSAX  observation in 1998  (Iaria et  al., 2001),
appeared by the time of a RXTE observation in 2001 May, but were again
not present in 2001 December (Smale et al. 2002).

Iaria et al.  (2001) fitted  the 0.1--100 keV BeppoSAX spectrum of XB
1254--690   adopting  a  multicolor   disk-blackbody  with   an  inner
temperature of 0.85 keV plus  a Comptonized component with an electron
temperature of 2 keV and an  optical depth of 19. The authors detected
a hard excess  at around 20 keV which might be  accounted for adding a
bremsstrahlung model with a temperature  of 20 keV. Finally, there was
also evidence  in the BeppoSAX  observation for an absorption  edge at
1.27 keV with an optical depth of 0.15.  Iaria et al.  (2001) proposed
that the  Comptonized component could originate in  a spherical cloud,
or   boundary  layer,   surrounding   the  neutron   star  while   the
bremsstrahlung component probably  originates in an extended accretion
disk corona with a radius of  $10^{10}$ cm and a mean electron density
of $1.7 \times 10^{14}$ cm$^{-3}$.

Boirin \& Parmar (2003) studied XB 1254--690 using two XMM-Newton EPIC
pn observations  in timing  and small window  mode with  an exposure
time of 10  and 13 ks, respectively.  The  continuum of the persistent
emission  in the  0.6--10  keV EPIC  pn  spectrum was  fitted using  a
multicolor disk-blackbody, with an inner temperature between 2 and 2.3
keV, plus  a power-law component with  a photon index  between 2.2 and
2.3.   Furthermore Boirin  \&  Parmar (2003)  detected a  \ion{Fe}{xxvi}
K$_\alpha$ and  a \ion{Fe}{xxvi}  K$_\beta$ absorption line  centered at
6.96 and 8.16 keV, with upper limits on the line widths between 95 and
120 eV, and between 80 and 170 eV, respectively; finally they observed
a soft  excess around 1  keV that was  modeled by a  Gaussian emission
line  centered at  0.93  keV,  with a  line width  of  175  eV and  an
equivalent width of $\sim 30$ eV.

Diaz  Trigo et  al.   (2005) modeled  the  changes in  both the  X-ray
continuum and the Fe absorption  features during dipping of six bright
LMXBs observed  by XMM-Newton, included XB  1254--690.  They concluded
that the dips were produced by an increase in the column density and a
decrease  in the  ionization  state of  the  highly ionized  absorber.
Moreover, outside of the dips,  the absorption line properties did not
vary strongly with orbital phase, this implied that the ionized plasma
had a cylindrical geometry with  a maximum column density close to the
plane of the  accretion disk.  Since dipping sources  are normal LMXBs
viewed  from close  to the  orbital  plane this  implied that  ionized
plasmas are a  common feature of LMXBs.  In  particular, Diaz Trigo et
al.  (2005)  included absorption from  a photoionized plasma  (xabs in
SPEX) in the spectral model to  account for the narrow features near 7
keV.  Reanalysing the  XMM data  of XB  1254--690 described  above the
authors obtained  a ionization parameter associated  to the persistent
emission of log($\xi$)$=4.3 \pm 0.1$ erg cm$^{-2}$ s$^{-1}$.
 
Finally we note that  the improved sensitivity and spectral resolution
of Chandra  and XMM-Newton are  allowing to observe  narrow absorption
features, from highly  ionized ions (H-like and He-like),  in a larger
and  larger number of  X-ray binaries.   These features  were detected
from the  micro-quasars GRO J1655--40  (Ueda et al.  1998;  Yamaoka et
al.  2001) and  GRS 1915+105 (Kotani et al.  2000;  Lee et al.  2002).
Recently  the Chandra  High-Energy  Transmission Grating  Spectrometer
(HETGS) observations  of the black hole candidate  H~1743--322 (Miller
et al.  2004) have  revealed the presence of blue-shifted \ion{Fe}{xxv}
and  \ion{Fe}{xxvi} absorption  features  suggesting the  presence of  a
highly-ionized  outflow.    All  LMXBs  that   exhibit  narrow  X-ray
absorption  features are  all known  dipping sources  (see Table  5 of
Boirin  et al.   2004)  except for  GX  13+1. This  source shows  deep
blue-shifted  Fe  absorption features  in  its  HETGS spectrum,  again
indicative of outflowing material (Ueda et al. 2004). We conclude this
brief resume observing that a  recent Chandra spectral analysis of the
dipping source XB 1916--053  showed the presence of several absorption
features  associated to  \ion{Ne}{x}, \ion{Mg}{xii},  \ion{Si}{xiv} and
\ion{S}{xvi}, \ion{Fe}{xxv}  and \ion{Fe}{xxvi}  (see Iaria et  al., 2006;
Juett \& Chakrabarty, 2006).

In this work we present  a Chandra spectral analysis of the persistent
emission  from the  dipping source  XB  1254--690 in  the 0.7--10  keV
energy range using a long  53 ks Chandra observation.  The observation
covered around 3.8  orbital periods and the lightcurve  showed no dips
at  all.  We  detected  a \ion{Fe}{xxvi}  K$_\alpha$ absorption  line,
already  observed   with  the  XMM  observation;   the  better  energy
resolution  of  Chandra and  the  larger  statistics  allowed to  well
determine its  width.  We  discuss that the  \ion{Fe}{xxvi} absorption
line  was  produced in  a  photoionized  absorber  placed between  the
coronal radius and the outer edge of the accretion disk.

\section{Observation} 

XB 1254--690 was observed with the Chandra observatory on 2003 October
10 from 01:58:44 to 17:15:55  UT using the HETGS.  The observation had
a total integration  time of 53 ks, and was  performed in timed graded
mode.  The HETGS  consists of two types of  transmission gratings, the
Medium Energy  Grating (MEG) and  the High Energy Grating  (HEG).  The
HETGS  affords  high-resolution  spectroscopy  from  1.2  to  31  \AA\
(0.4--10  keV)  with a  peak  spectral  resolution of  $\lambda/\Delta
\lambda  \sim 1000$ at  12 \AA\  for HEG  first order.   The dispersed
spectra  were recorded  with an  array of  six  charge-coupled devices
(CCDs)  which are  part  of the  Advanced  CCD Imaging  Spectrometer-S
(Garmire             et             al.,            2003)\footnote{See
  http://asc.harvard.edu/cdo/about\_chandra  for  more details.}.   We
processed the  event list using available software  (FTOOLS v6.0.2 and
CIAO v3.2  packages) and  computed aspect-corrected exposure  maps for
each spectrum, allowing  us to correct for effects  from the effective
area of the CCD spectrometer.

The brightness  of the source required additional  efforts to mitigate
``photon pileup'' effects. A 512  row ``subarray'' (with the first row
= 1) was applied during the observation reducing the CCD frame time to
1.74 s.   Pileup distorts the  count spectrum because  detected events
overlap  and  their  deposited  charges  are  collected  into  single,
apparently more  energetic, events.   Moreover, many events  ($\sim 90
\%$) are  lost as the grades of  the piled up events  overlap those of
highly energetic background particles and  are thus rejected by the on
board software.  We, therefore, ignored the zeroth-order events in our
spectral analysis.  On  the other hand, the grating  spectra were not,
or only  moderately (less  than 10 \%),  affected by pileup.   In this
work we  analysed the 1st-order HEG  and MEG spectra; since  a 512 row
subarray  was applied  the 1st-order  HEG  and MEG  energy range  were
shrinked to 0.9--10 keV and 0.7--7 keV, respectively.
\begin{figure}
\resizebox{\hsize}{!}{\includegraphics{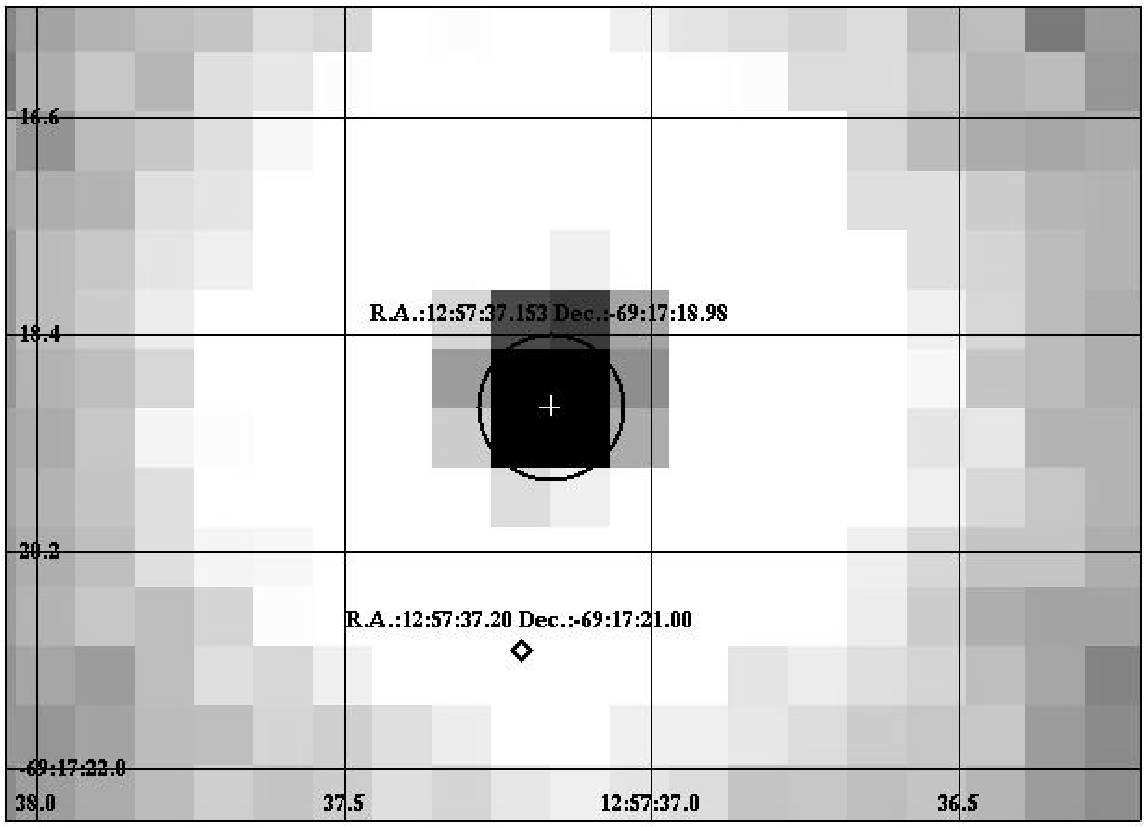}}
\caption{Region  of sky  around  the Chandra  zero-order  image of  XB
  1254--690. The  coordinate system is referred to J2000.   The black
  circle centered  to the best  estimation (white cross point)  of the
  Chandra  position has  a  radius of  0.6\arcsec.  The diamond  point
  indicates  the position  of  XB 1254--690  (J2000)  reported by  the
  Heasarch  ``Coordinate  Converter'' tool  (see  text).  The  angular
  separation between the two points  is 2\arcsec.}
\label{fig0}
\end{figure}

To  determine the  zero-point position  in the  image as  precisely as
possible,  we estimated the  mean crossing  point of  the zeroth-order
readout trace  and the tracks  of the dispersed  HEG and MEG  arms. We
obtained    the   following    coordinates:   R.A.=$12^h57^m37^s.153$,
Dec.=$-69^{\circ}  17\arcmin  18\arcsec.98$   (J2000.0,  with  a  90\%
uncertainty circle of the absolute position of 0.6\arcsec\footnote{See
  http://cxc.harvard.edu/cal/ASPECT/celmon/  for more  details.}).  We
compared the  coordinates of XB  1254--690 obtained by us  to the
coordinates  of the  source reported  by  the the  NASA HEASARCH  tool
``Coordinate                                   Converter''\footnote{See
  http://heasarch.gsfc.nasa.gov/cgi-bin/Tools/convcoord/convcoord.pl}.
The coordinates  obtained with the tool  were R.A.= $12^h57^m37^s.20$,
Dec.=-69$^{\circ}$17\arcmin 21\arcsec.0  (referred to J2000.0), having
an angular  separation from the  Chandra coordinates of  2\arcsec.  We
noted that  Boirin \&  Parmar (2003) could  not estimate  more precise
values of  the coordinates because  the spatial resolution of  all the
XMM-Newton EPIC  cameras is larger than  2\arcsec.  In Fig.~\ref{fig0}
we reported the region of sky around the Chandra zero-order image of 
XB 1254--690.
\begin{figure}
\resizebox{\hsize}{!}{\includegraphics{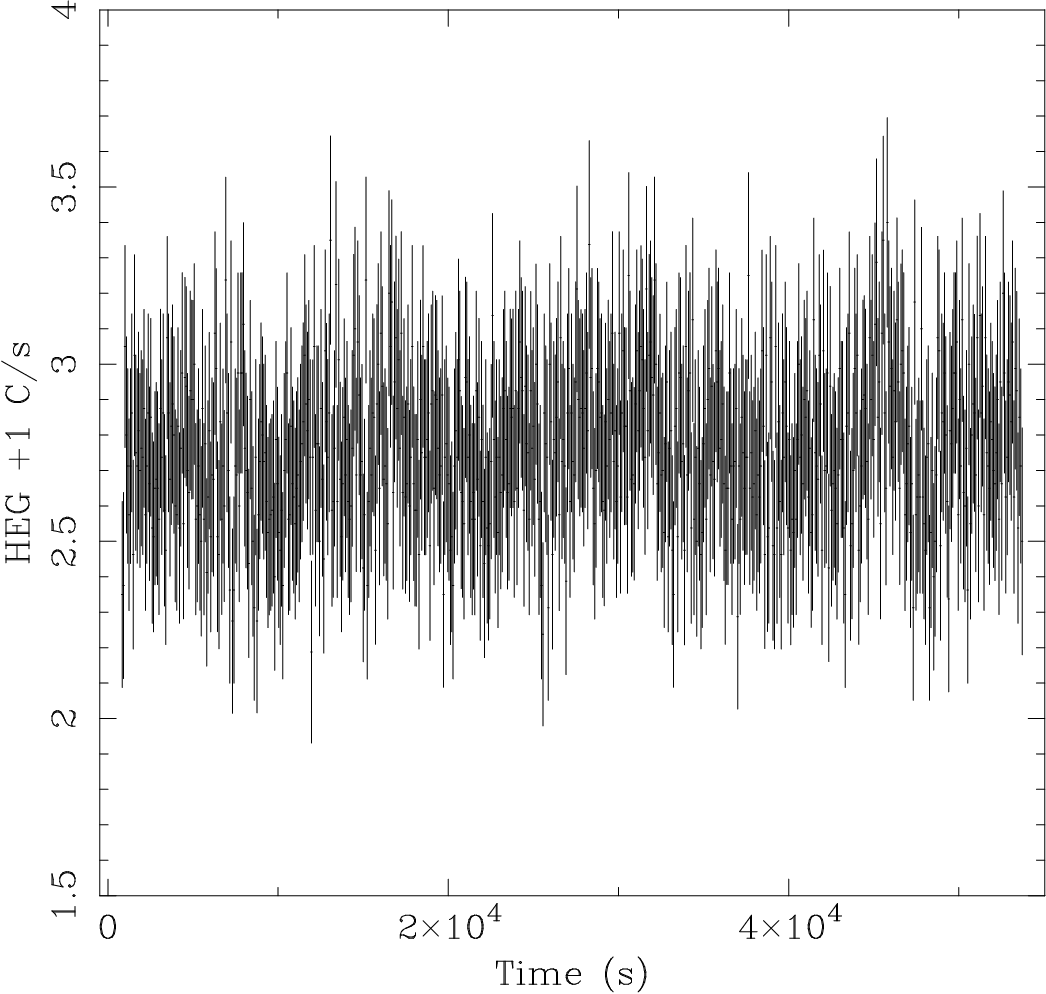}}
\caption{The 53 ks lightcurve  of XB 1254--690. During the observation
  neither bursts  or dips were observed.  The  used events corresponds
  to the positive first-order HEG.  The bin time is 80 s.}
\label{fig1}
\end{figure}
 
In Fig.~\ref{fig1} we showed the  80 s bin time lightcurve taking into
account only  the events  in the positive  first-order HEG.   The mean
count rate in  the persistent state was 2.8  count s$^{-1}$.  During the
observation neither bursts or dips  were observed.

\section{Spectral Analysis}
\begin{table*} 
\caption{ \footnotesize \linespread{1} 
 The  photoelectric absorption is indicated as
  N$_H$.   Uncertainties are  at 90\%  confidence level  for  a single
  parameter; upper limits are at 95\% confidence level.  kT$_{in}$ and
  $k$   are,  respectively,   the  accretion   disk   temperature  and
  normalization  in units  of $R_{in}^2/D^2_{10}  \cos  \theta$, where
  $R_{in}$  is  the inner  radius  of the  accretion  disk  in km  and
  D$_{10}$ is the distance in units of 10 kpc.  kT$_{BB}$ and N$_{BB}$
  are,  respectively, the blackbody  temperature and  normalization in
  units of  L$_{39} / D^2_{10}$,  where L$_{39}$ is the  luminosity in
  units of  10$^{39}$ ergs s$^{-1}$; kT$_e$,  $\tau$, and N$_{CompST}$
  indicate  the  electron  temperature,  the optical  depth,  and  the
  normalization of  the CompST model of the  Comptonizing cloud around
  the  neutron  star,   respectively;  N$_{po}$  and  N$_{po}$  cutoff
  indicate the  normalization of  the power law  and the  cutoff power
  law,  in unit  of photons  keV$^{-1}$ s$^{-1}$  cm$^{-2}$ at  1 keV.
  E$_{edge}$ and  $\tau_{edge}$ indicate the threshold  energy and the
  optical depth associated to the absorption edge.}
\label{tab0}
\centering
 \begin{tabular}{l c c c c }
\hline
\hline
  &   MCD  & MCD  &   Blackbody   \\
  &     + &   +   &     +           \\ 
  & power law &CompST& Power Law Cutoff \\
\hline
Parameters &  & & & \\
\hline

$N_{\rm H}$ $\rm (\times 10^{22}\;cm^{-2})$ 
& $0.240^{+0.013}_{-0.017}$ 
& $0.199^{+0.023}_{-0.010}$ 
& $0.175^{+0.014}_{-0.016}$ \\
&\\

E$_{edge}$ (keV)
&     --
&   1.276 (fixed)
&     --
 \\  

$\tau_{edge}$ 
&      --
&   $<0.014$ 
&      -- \\
&\\

kT$_{in}$ (keV)
&    $2.470 \pm 0.020$   
&   0.855 (fixed) 
&      -- \\

$k$
&  $1.393^{+0.044}_{-0.065}$ 
&  $12.2^{+6.7}_{-5.7}$ 
&      -- \\
&\\

kT$_{BB}$ (keV)
&    --   
&      --
&  $1.466^{+0.060}_{-0.115}$      \\

N$_{BB}$ ($\times 10^{-3}$)
&   --
&  -- 
&   $2.37^{+0.73}_{-0.87}$     \\
&\\

photon index 
&  $2.31^{+0.21}_{-0.20}$ 
&    -- 
&    -- \\  
 
N$_{po}$
&  $0.0700 \pm 0.0035$ 
&    -- 
&    -- \\
&\\

photon index 
&  -- 
&    -- 
&   $1.033^{+0.069}_{-0.077}$ \\  
 
N$_{po}$ cutoff 
&  --
&    -- 
&    $0.1453^{+0.0040}_{-0.0047}$ \\
 
E$_{co}$ (keV)
&  --
&  -- 
&   5.9 (fixed) \\
&\\

kT$_e$ (keV)
&     --
& $1.759^{+0.046}_{-0.039}$   
&     --\\
  
$\tau$ 
&     --
& $24.6^{+2.8}_{-2.0}$  
&     -- \\

N$_{CompST}$  
&     --
& $0.103^{+0.021}_{-0.023}$   
&     --\\
&\\
 
L$_{0.6-10 {\rm keV}}$ erg s$^{-1}$  
&     $1.2 \times 10^{37}$
&   $1.2 \times 10^{37}$  
&   $1.2 \times 10^{37}$  \\
&\\
 
$\chi^2$(d.o.f.)
&  2406(2723) 
&  2447(2722)  
&  2412(2723)  \\
\hline
\hline
\end{tabular}
\end{table*}
  
Because the  lightcurve and the hardness  ratio of the  source did not
show changes  during the whole  observation we selected  the 1st-order
spectra from the HETGS data  along the entire observation with a total
exposure time of  52 ks.  Data were extracted  from regions around the
grating arms; to avoid overlapping between HEG and MEG data, we used a
region size  of 25 and  33 pixels for  the HEG and  MEG, respectively,
along  the cross-dispersion  direction.  The  background  spectra were
computed, as usual,  by extracting data above and  below the dispersed
flux.  The contribution  from the background is $0.4  \%$ of the total
count  rate.  We  used  the  standard CIAO  tools  to create  detector
response files  (Davis 2001) for the HEG  -1 (MEG -1) and  HEG +1 (MEG
+1) order  (background-subtracted) spectra.  After  verifying that the
negative and  positive orders were  compatible with each other  in the
whole   energy  range   we  coadded   them  using   the   script  {\it
  add\_grating\_spectra} in the CIAO software, obtaining the 1st-order
MEG spectrum and the 1st-order  HEG spectrum.  Finally we rebinned the
resulting 1st-order  MEG and 1st-order  HEG spectra to 0.015  \AA\ and
0.0075  \AA, respectively.

To fit  the continuum we used  the rebinned spectra in  the 0.7--7 keV
and 0.9--10 keV for first-order MEG and first-order HEG, respectively.
Initially we tried  to fit the continuum using  a single component and
adding a  systematic edge at around  2.07 keV with  a negative optical
depth of  $\sim -0.2$ to take  in account of  an instrumental artifact
(see Miller et al.  2002,  and references therein). Models composed of
an absorbed power-law or an  absorbed multicolor disk (MCD; Mitsuda et
al.,  1984)  gave  $\chi^2$(d.o.f.)   of  3483(2725)  and  2733(2725),
respectively, and the data showed large residuals along the whole used
energy range indicating  a bad choice of the  continuum.  Following we
tried  models  composed  of  two  absorbed  components  finding  three
statistically equivalent  models giving good fits,  the first composed
of a  power law plus  a MCD component  as adopted by Boirin  \& Parmar
(2003),  the second  composed of  a MCD  component plus  a Comptonized
component (CompST  in XSPEC) as adopted  by Iaria et  al.  (2001), and
the third  composed of  a blackbody  plus a power  law with  cutoff as
reported by Smale  et al.  (2002); we obtained  a $\chi^2$(d.o.f.)  of
2406(2723), 2447(2722), and 2412(2723), respectively.

From the fit of the model composed of a MCD plus a power law component
we found an absorbing equivalent hydrogen column of N$_H = 0.24 \times
10^{22}$ cm$^{-2}$,  a photon index of 2.3,  a power-law normalization
of  0.07,  a  temperature  of   the  MCD  component  of  2.47  keV,  a
normalization  associated  to  the  MCD  of 1.39,  and  an  unabsorbed
luminosity, assuming a  distance to the source of  10 kpc (Courvoisier
et  al., 1986;  hereafter  we assume  this  value of  distance to  the
source), of $\sim 1.2 \times 10^{37}$ ergs s$^{-1}$ in the 0.6--10 keV
energy range.  The best-fit  values and the unabsorbed luminosity were
comparable to  those obtained by  Boirin \& Parmar  (2003), indicating
that  the  source  was  in   a  similar  spectral  state  during  this
observation and the previous XMM observations.  To confirm this result
we analysed the RXTE ASM lightcurve of XB 1254--690.  We observed that
during the  two XMM  observations the  ASM count rate  was $  2.24 \pm
0.16$ and $2.02  \pm 0.26$ C/s, while during  this observation the ASM
count rate  was $2.66 \pm  0.30$ C/s with  a similar intensity  of the
source as obtained from the spectral analysis.

From  the fit  of  the model  composed  of a  MCD  plus a  Comptonized
component (CompST in XSPEC)  we found an absorbing equivalent hydrogen
column  of N$_H  \simeq 0.20  \times 10^{22}$  cm$^{-2}$,  an electron
temperature kT$_e$ of 1.76 keV,  an optical depth $\tau$ associated to
the  Comptonizing region of  25, and  a normalization  N$_{compST}$ of
0.10.   We  fixed  the  internal  temperature of  the  accretion  disk
kT$_{in}$ to 0.855  keV as reported by Iaria et  al.  (2001) finding a
MCD normalization $ k$ of  12.2.  We found an unabsorbed luminosity of
$\sim 1.2  \times 10^{37}$  ergs s$^{-1}$ in  the 0.6--10  keV energy,
compatible to that obtained by Iaria et al.  (2001) in the same energy
range.  This  was also confirmed looking  at the ASM  lightcurve of XB
1254--690,  during the  BeppoSAX observation  the ASM  count  rate was
$1.96 \pm 0.60$ C/s (see Fig.~\ref{fig2b}).   Iaria  et al.  (2001)
detected  an absorption  edge at  1.276 in  the BeppoSAX  spectrum, we
added  this  component  fixing  the  energy threshold  and  finding  a
corresponding  optical  depth   $<0.014$,  not  compatible  to  that
previously detected of $0.154 \pm 0.059$.

From the  fit of the  model composed of  a blackbody plus a  power law
with cutoff we  found an absorbing equivalent hydrogen  column of N$_H
\simeq  0.18  \times   10^{22}$  cm$^{-2}$,  a  blackbody  temperature
kT$_{BB}$ of 1.47 keV, and  a blackbody normalization N$_{BB}$ of $2.3
\times 10^{-3}$.  We  fixed the energy cutoff of the  power law to 5.9
keV, as reported  by Smale et al.  (2002)  for the persistent emission
of XB 1254--690, finding a photon index of 1.03 and a
power-law  normalization  of  0.145.   All  the  best-fit  values  are
compatible to those obtained by Smale et al.  (2002) except N$_{BB}$
that  was slightly  lower  indicating a  lower  emission of  blackbody
during the  Chandra observation, however  we noted that the  ASM count
rate during  the XTE observation  was $2.46 \pm 0.30$  C/s, compatible
with the ASM count rate during the observation discussed in this work.
 
\begin{figure}
\resizebox{\hsize}{!}{\includegraphics{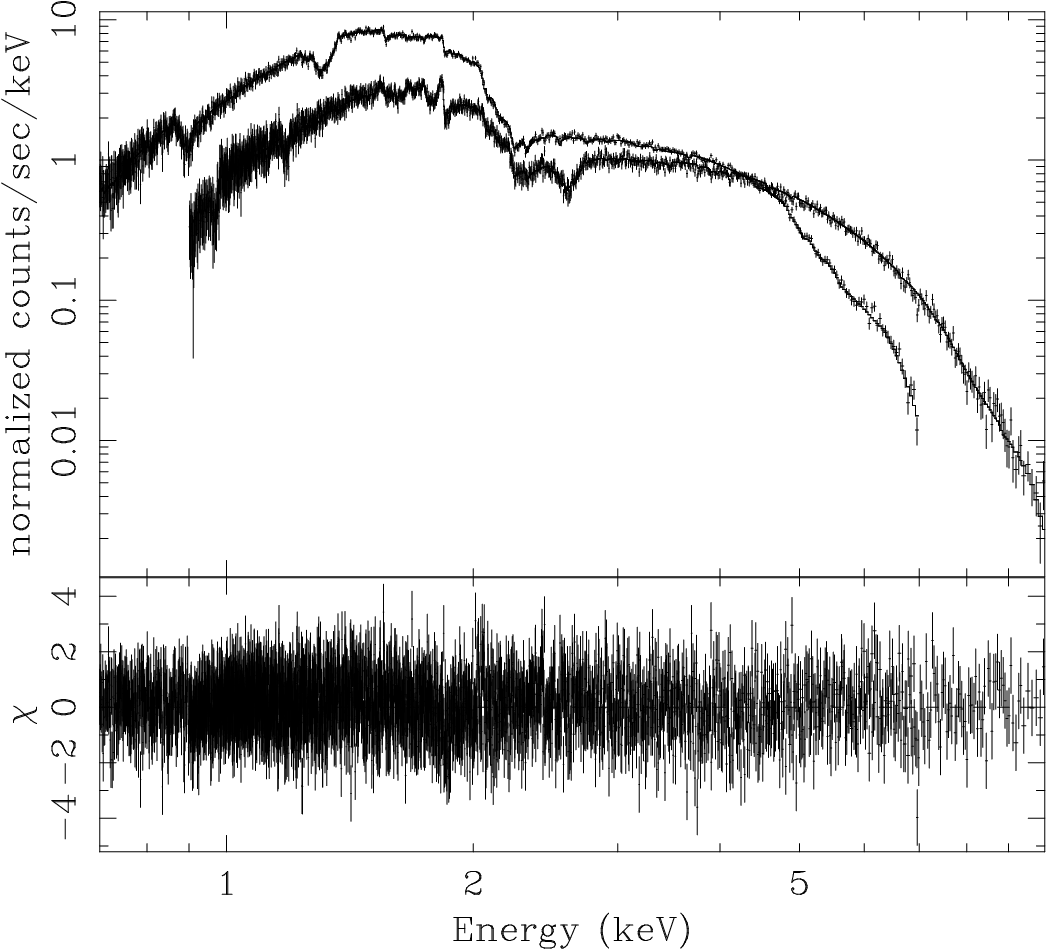}}
\caption{Data and  residuals of the persistent emission  in the energy
  range 0.7--10  keV for  the rebinned 1st-order  MEG and  HEG spectra
  (see text).  The continuum is fitted by an absorbed power law plus a
  MCD (Mitsuda et al., 1984) component.  From the residuals is evident
  an absorption feature at around 7 keV.  }
\label{fig2}
\end{figure}
\begin{figure}
  \resizebox{\hsize}{!}{\includegraphics{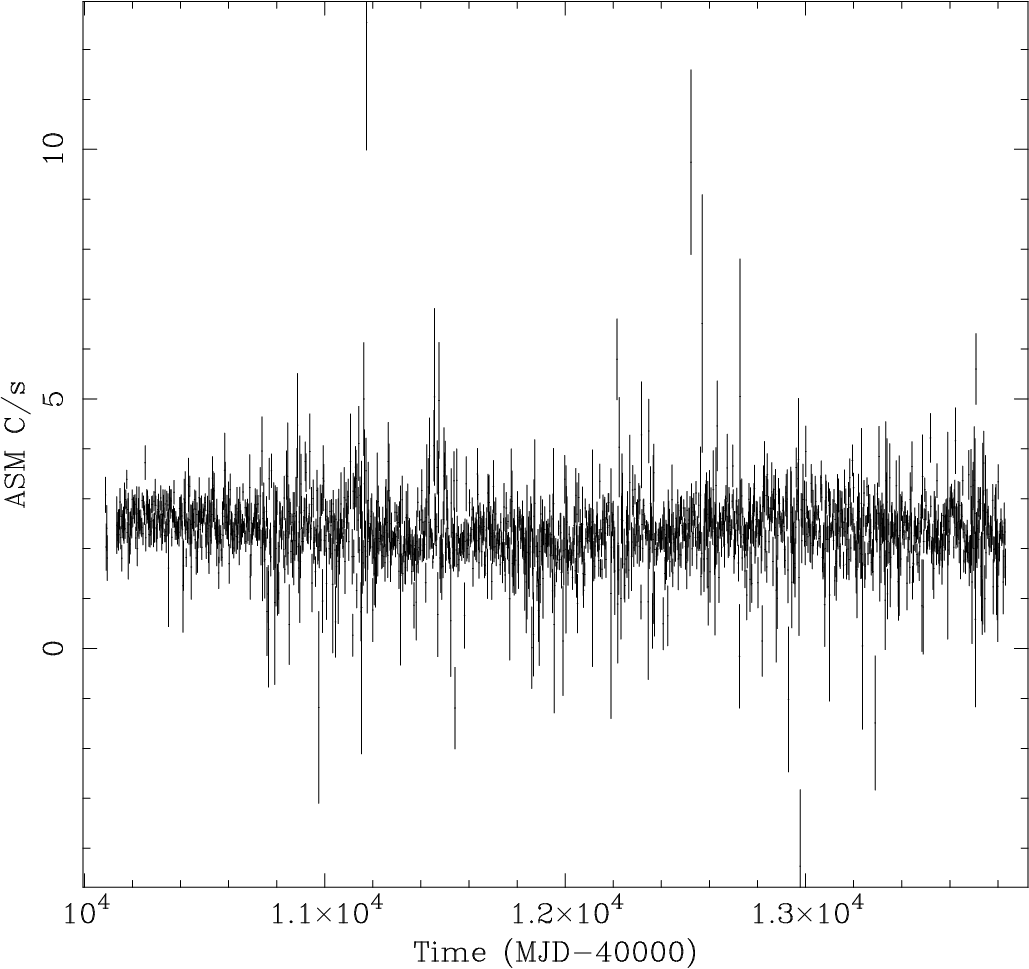}
    \includegraphics{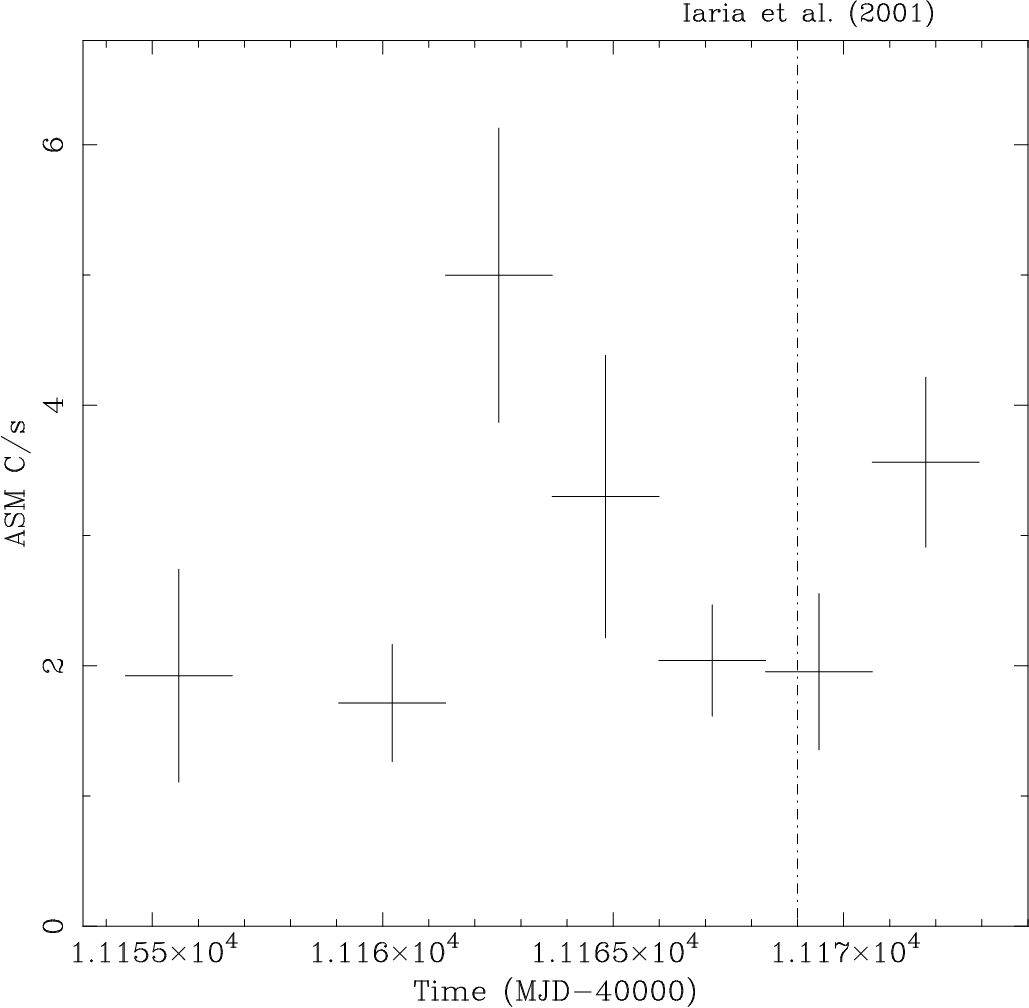}\includegraphics{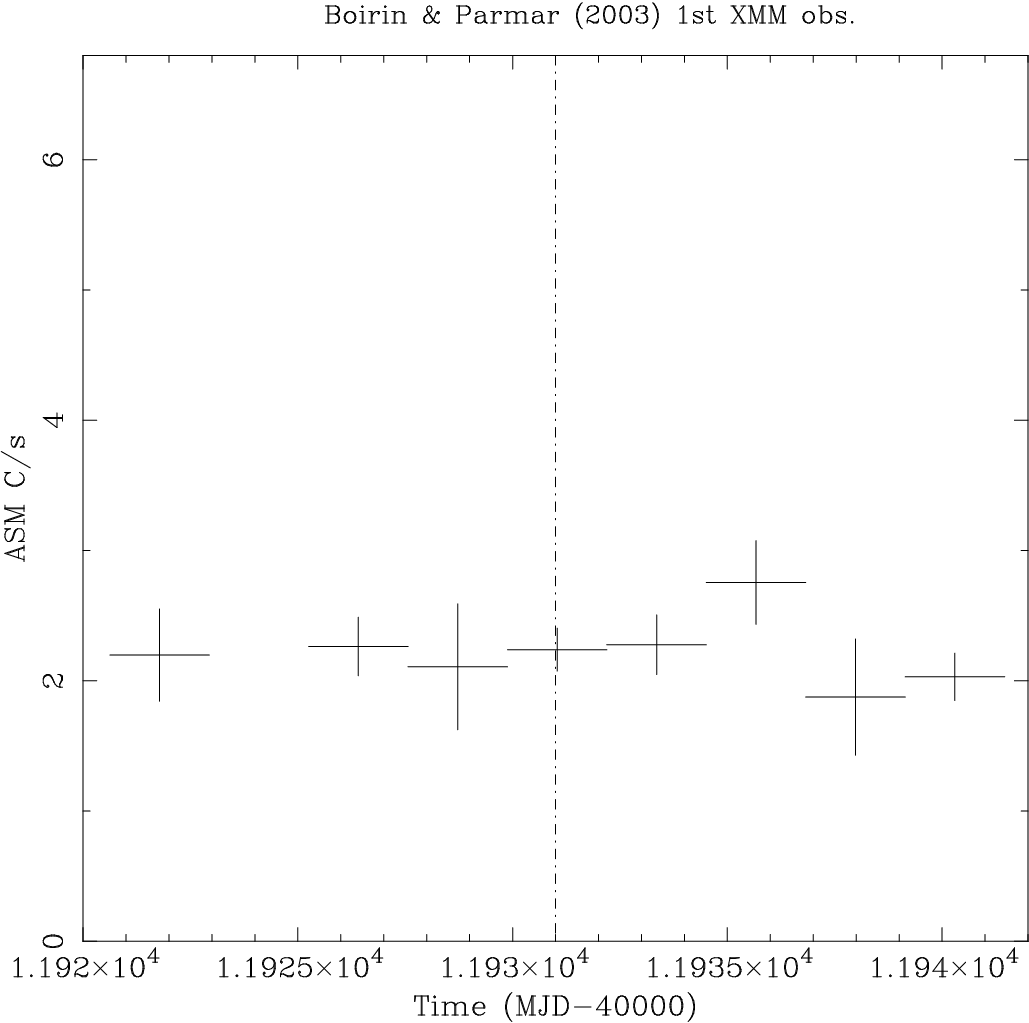}}\\
   \resizebox{\hsize}{!}{                   \includegraphics{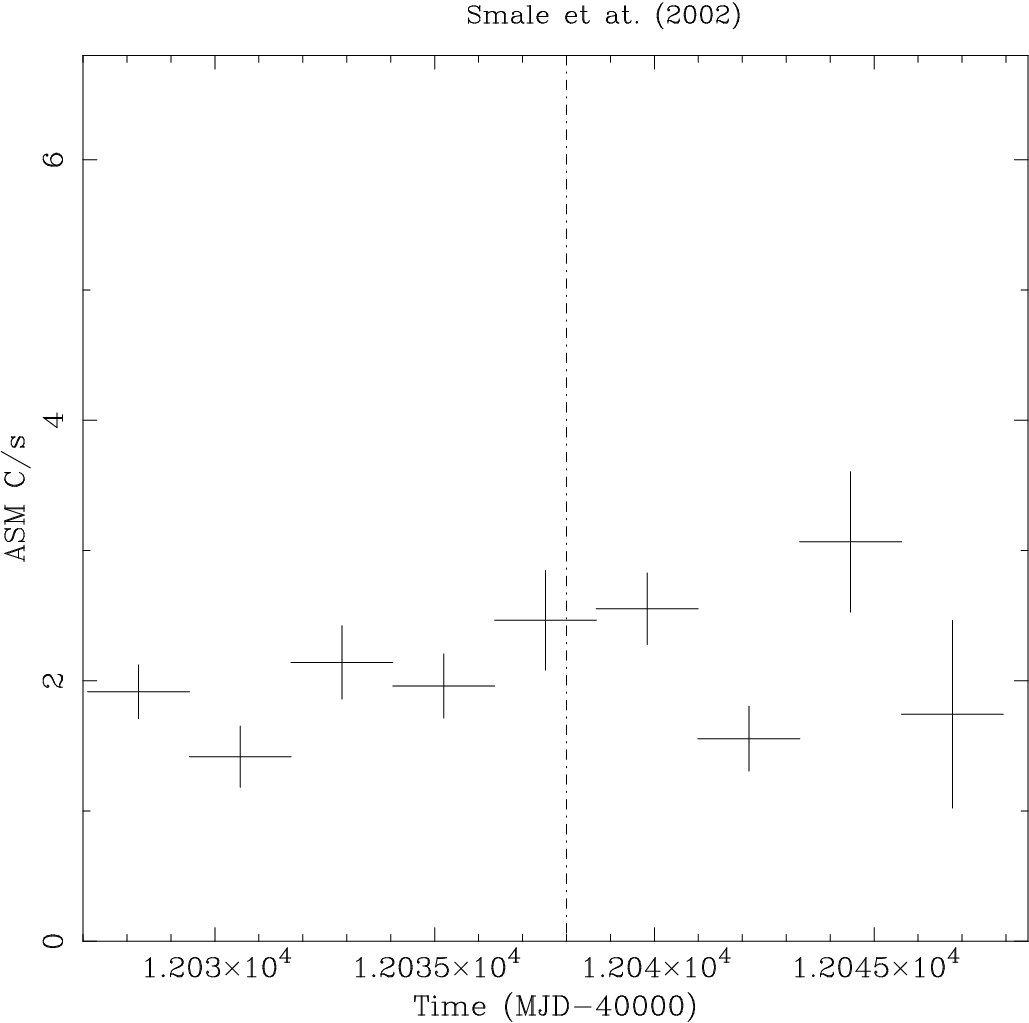}
    \includegraphics{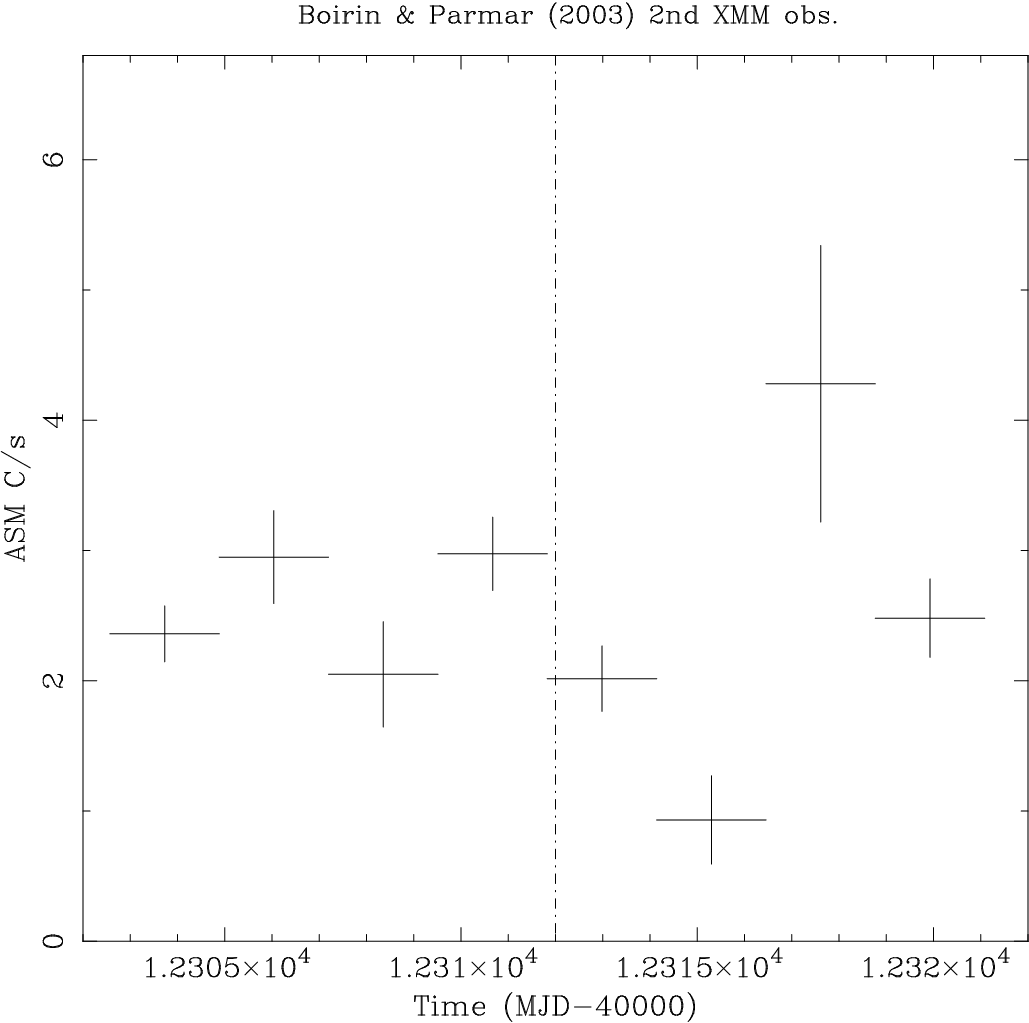} \includegraphics{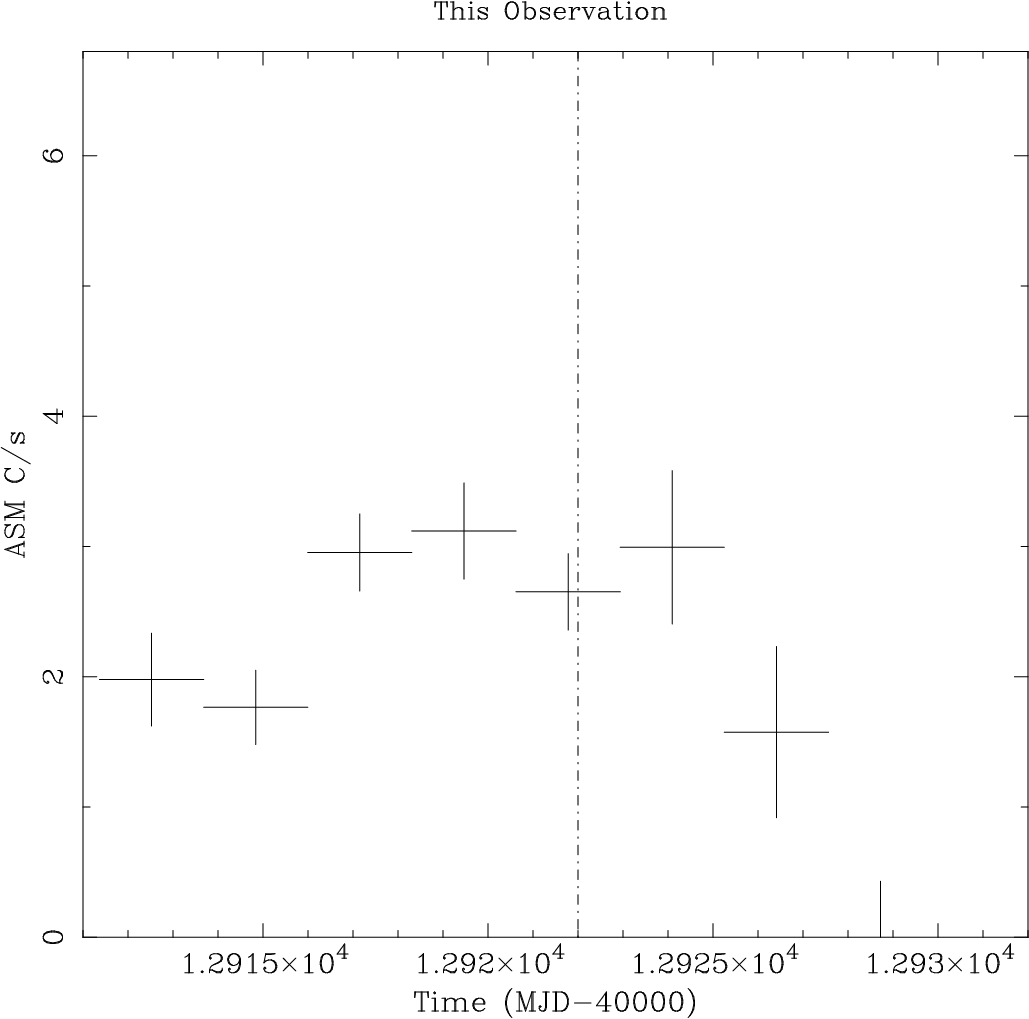}}
  \caption{The RXTE  ASM lightcurve of XB  1254--690 (top-left panel).
    In the  others panels  we expanded the  temporal scale  to clearly
    show the ASM count rate during the previous observations
    of the source.  The dash-pointed vertical lines indicate the start
    times   of  the  BeppoSAX   observation  (top-center   panel),  XMM
    observations  (top-right  and   bottom-center  panels),  RXTE  PCA
    observation    (bottom-left   panel),    and    this   observation
    (bottom-right panel). }
\label{fig2b}
\end{figure}
 
Finally we note  that the best-fit values of  the absorbing equivalent
hydrogen column  N$_H$ of  each model were  smaller than  $0.29 \times
10^{22}$  cm$^{-2}$, the  Galactic column  density measured  toward XB
1254--690 in the  21 cm radio survey of Dickey  \& Loekman (1990).  We
reported the  parameters of  the continuum models  in Tab.~\ref{tab0},
since the three  adopted models gave similar residuals  we adopted the
continuum model  composed of  MCD plus power  law in the  following of
this  section.  In  the next  section we  briefly discussed  the three
models.  We reported the data and the residuals in Fig.~\ref{fig2} and
the ASM lightcurve of XB 1254--690 in Fig.~\ref{fig2b}.

\begin{figure}
\resizebox{\hsize}{!}{\includegraphics{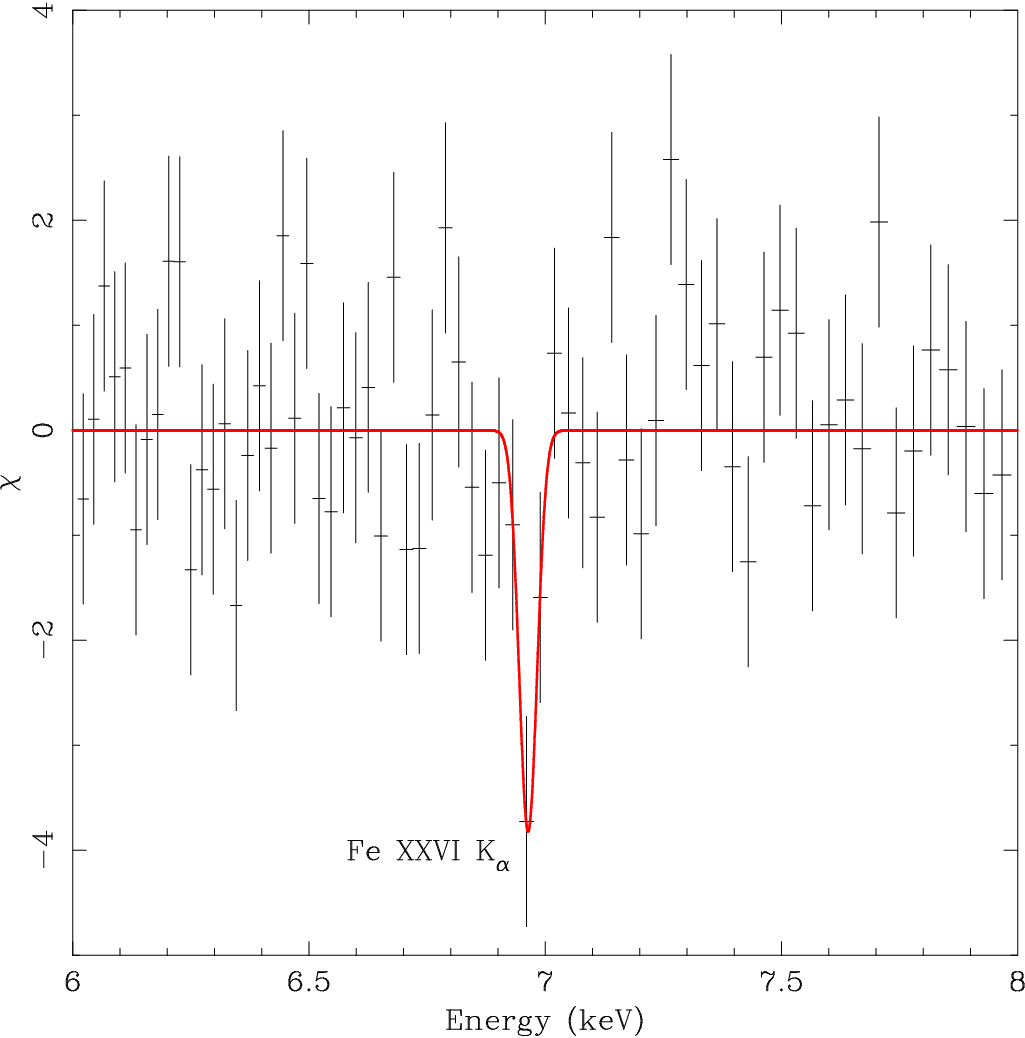}}
   \caption{Residuals  with  respect  to  the  best-fit  model  of  the
    continuum  reported  in  Tab.~\ref{tab1}   of the  1st-order HEG 
  spectrum.  In the  panel the residuals are plotted in
    the energy  intervals around the \ion{Fe}{xxvi}  absorption line.}
\label{fig3}
\end{figure}

From an accurate analysis of  the residuals we detected two absorption
features between  0.8 and  0.9 keV, and  an absorption feature  near 7
keV, their presence was  model independent.  To resolve the absorption
features  we  fixed the  continuum  and  added  a Gaussian  line  with
negative normalization  for each feature.   We used the  1st-order MEG
spectrum  to resolve  the  absorption  features below  1  keV and  the
1st-order HEG spectrum  to resolve that near 7  keV.   The
   absorption features  below 1 keV  were  centered at 0.835 and
  0.849 keV with  a significance of 3$\sigma$ and  2.4$\sigma$, and 
  equivalent  widths of  -0.65, and  -0.69 eV,  respectively.  Since
  these lines have a low significance and a no clear identification we
  retained that they could be  due to  statistical fluctuations in the
  MEG instrument (see the Discussion).
 In  the  6--8  keV energy  range  we detected  an
absorption  line centered  at 6.962  keV  with a  significance of
  3.7$\sigma$,   corresponding  to  \ion{Fe}{xxvi}   K$_\alpha$;  the
equivalent width  was -15.8 eV.   In Fig.~\ref{fig3}  we
showed the  residuals with  respect to the  continuum in the  6--8 keV
energy range, in Table~\ref{tab1}  we reported the best-fit parameters
of the three observed absorption lines.

\begin{table} 
\caption{ \footnotesize \linespread{1}  
The parameters of the Gaussian absorption  lines
  are E, $\sigma$, I, EW, and FWHM indicating the centroid, the width,
  the intensity    in units of photons  s$^{-1}$ cm$^{-2}$,
  the equivalent width,  and the Full Width Half  Maximum of the lines,
  respectively.}
\label{tab1}
\centering
 \begin{tabular}{l c }
\hline
\hline
& Parmeters \\ 
 &  \\
\hline

E$$ (keV)
& $0.83505^{+0.00051}_{-0.00055}$ \\

$\sigma$ (eV)
&  $<1.2$ \\

I$$ ($\times 10^{-4}$ cm$^{-2}$ s$^{-1}$)
& $-1.11^{+0.57}_{-0.83}$ \\

EW$$ (eV)
&  $-0.65^{+0.33}_{-0.48}$ \\

FWHM$$ (km s$^{-1}$) 
&  $< 1020$ \\
 
&\\

 &  \\
\hline

E$$ (keV)
& $0.8486^{+0.0027}_{-0.0014}$ \\

$\sigma$ (eV)
&  $1.00^{+4.09}_{-0.73}$ \\

I$$ ($\times 10^{-4}$ cm$^{-2}$ s$^{-1}$)
& $-1.16^{+0.76}_{-1.60}$ \\

EW$$ (eV)
&  $-0.69^{+0.45}_{-0.95}$ \\

FWHM$$ (km s$^{-1}$)
&  $830^{+3400}_{-670}$ \\
 &\\

\ion{Fe}{xxvi} K$_{\alpha}$   &  \\
\hline

E$$  (keV)
& $6.962^{+0.012}_{-0.015}$  \\

$\sigma$ (eV)
&  $<35$ \\

I$$ ($\times 10^{-4}$ cm$^{-2}$ s$^{-1}$)
&  $-1.22^{+0.52}_{-0.66}$ \\

EW$$ (eV)
& $-15.8^{+6.7}_{-8.5}$  \\
FWHM$$ (km s$^{-1}$)
&  $<3550$ \\
\hline
\hline
\end{tabular}
\end{table}

\section{Discussion}
We  have  analyzed a  53  ks  Chandra  observation of  the  persistent
emission from XB~1254--690. The  position of the zeroth-order image of
the  source provides  improved  X-ray coordinates  for XB~1254--690  
(R.A.=$12^h57^m37^s.153$,  Dec.=$-69^{\circ}  17\arcmin 18\arcsec.98$),
with an angular separation of 2\arcsec\ to the X-ray position reported
by the  ``Coordinate Converter'' tool  (see section 2),  impossible to
distinguish  from the analysis  of the  two previous  XMM observations
(Boirin \& Parmar, 2003) because  the larger spatial resolution of all
the  XMM  EPIC cameras.   We  performed  a  spectral analysis  of  the
persistent  emission using  the 1st-order  MEG and  HEG  spectra.  The
continuum emission  was well fitted  by three different models:  a MCD
component  plus  a power  law,  a  MCD  component plus  a  Comptonized
component (CompST  in XSPEC),  and a blackbody  plus a power  law with
cutoff.  We obtained from these models an unabsorbed luminosity of
$1.2 \times  10^{37}$ erg  s$^{-1}$ in the  0.6--10 keV  energy range.
The first model was the same adopted by Boirin \& Parmar (2003) to fit
the XMM  spectra of  XB~1254--690 giving similar  best-fit parameters,
the second model  was the same adopted by Iaria et  al.  (2001) in the
0.6--10 keV  energy range using BeppoSAX data  finding, again, similar
best-fit  parameters,  and, finally,  the  third  model  was the  same
adopted by Smale  et al.  (2002) using RXTE PCA data  and also in this
case  we  obtained  comparable  best-fit  values.   The  statistically
equivalence of  the three  models suggested that  the state of  the XB
1254--690 was stable during  the observations from 1998 December (date
of  the  BeppoSAX  observation)  up  to 2003  October  (date  of  this
observation).  This  conclusion was also suggested by  our analysis of
the  ASM lightcurve  that showed  a similar  value of  the  count rate
during the four observations.

We noted that  the model composed of a MCD component  plus a power law
appeared not  physically plausible, since  it was hard to  explain the
small inner  radius associated to  the best-fit parameters of  the MCD
component.  In fact, the inner radius associated to the temperature of
2.47 keV was R$_{in} \sqrt{\cos \theta} \simeq 1.2$ km and taking into
account  the  inclination angle  $\theta$  of  XB  1254--690, that  is
between 65$^{\circ}$ and 73$^{\circ}$ (Courvoisier et al.  1986, Motch
et  al.  1987),  we found  R$_{in} \simeq  2$ km,  much less  than the
neutron star  radius. On the  other hand the  model composed of  a MCD
component plus a Comptonized  component, already discussed by Iaria et
al. (2001),  implied that the inner  radius of the  accretion disk was
between 5  and 6 km.  Since  the electron scattering  could modify the
MCD emission (Shakura \& Sunyaev  1973; White, Stella \& Parmar 1988),
we corrected  the value  of the inner  radius and its  temperature for
this effect  obtaining R$_{in} \simeq 17$  km (see Iaria  et al., 2001
and  references   therein  for  a  more   detailed  discussion).   The
Comptonized  component  could  be  associated  to a  corona  having  a
temperature of 2 keV and  surrounding the neutron star, as proposed by
the Eastern Model (Mitsuda et al., 1989).  Finally, the model composed
of  a power  law with  energy cutoff  plus a  blackbody  component was
consistent to  the Birmingham Model (see  Church \& Balucinska-Church,
2004).  In  this case the  blackbody emission is  thermalized emission
from the neutron star and the cutoff power-law emission is produced in
an extended corona above the accretion disk.  Unfortunately the narrow
0.7--10 keV  energy range could  not allow us to  discriminate between
these models  although the scenario  proposed by the  Birmingham model
better described the line widths found in our analysis (see below).

At  high energies  we  detected a  \ion{Fe}{xxvi} K$_\alpha$  absorption
line.  The \ion{Fe}{xxvi} absorption line was already observed by Boirin
\& Parmar (2003)  using two XMM EPIC pn observations,  in that case an
upper limit of  95 eV was estimated on the line  width.  Thanks to the
higher spectral resolution  of Chandra HEG and to  an observation four
times longer, we found that the corresponding line width was $<35$ eV.
Furthermore,  Boirin  \&  Parmar  (2003) detected  a  weak  absorption
feature at $8.16 \pm  0.06$ keV proposing two possible interpretations
of  it: a) an  absorption line  associated to  \ion{Fe}{xxvi} K$_\beta$,
although  the  centroid  should have  to  be  8.26  keV and  then  not
compatible to the best-fit value reported by them, and b) a systematic
feature  due to  the uncertain  calibration of  the EPIC  pn,  in that
spectral region, at the time of their analysis.  We did not detect any
absorption feature which could be associated to \ion{Fe}{xxvi} K$_\beta$
but we could not exclude that  the low effective area of the 1st-order
HEG  above 7  keV  did  not allow  us  to detect  it.   To check  this
possibility we added a Gaussian  absorption line with the centroid and
the line width fixed to 8.16 keV and 10 eV, respectively.  We found an
upper limit of  the equivalent width $>-15$ eV  that was compatible to
that  obtained   from  the  XMM   observations,  then  we   could  not
discriminate between the two possible explanations about the origin of
this narrow feature given by Boirin \& Parmar (2003).

Since we  observed only the  \ion{Fe}{xxvi} K$_\alpha$ and did  not have
evidence  of  a \ion{Fe}{xxv}  absorption  lines  we  deduced that  the
ionization parameter  associated to this line was  log$(\xi)>4$ erg cm
s$^{-1}$.  We found the upper limit to the equivalent width associated
to the \ion{Fe}{xxv}  absorption line fixing its centroid  and its line
width  to 6.7002  keV and  10 eV,  respectively, and  finding  a value
$>-2.6$  eV.  We  estimated the  \ion{Fe}{xxv} and  \ion{Fe}{xxvi} column
densities  using the relation:  $$
\frac{W_\lambda}{\lambda}=\frac{\pi
  e^2}{m_e c^2}  N_j \lambda  f_{ij}=8.85 \times 10^{-13}  N_j \lambda
f_{ij} $$
where $N_j$ is  the column density for the relevant species,
$f_{ij}$  is the  oscillator strength,  $W_\lambda$ is  the equivalent
width  of the  line, and  $\lambda$ is  the wavelength  in centimeters
(Spitzer 1978, p. 52).  Adopting $f_{ij} = 0.798$ and $f_{ij} = 0.416$
for  \ion{Fe}{xxv}   and  \ion{Fe}{xxvi}  respectively   (see  Verner  \&
Yakovlev, 1995), and the  best parameters reported in Tab. \ref{tab1},
we  found N$_{\rm  Fe  XXV} \la  1.1  \times 10^{17}  $ cm$^{-2}$  and
N$_{\rm Fe XXVI} \simeq 3.5  \times 10^{17} $ cm$^{-2}$. We noted that
these values were similar to those  estimated by Lee et al. (2002) for
GRS  1915+105 where  the  ionization parameter  was  estimated by  the
authors to be  log($\xi$)$> 4.15$ erg cm s$^{-1}$.   A further confirm
about  the goodness  of  our  estimation of  $\xi$  could be  obtained
considering the Chandra spectroscopic results of the dipping source XB
1916--053 (Iaria et  al., 2006) which had an  unabsorbed luminosity of
$7.5 \times 10^{35}$ erg s$^{-1}$ in the 0.6--10 keV energy range.  In
that case, assuming an ionization  continuum consisting of a power law
with  $\Gamma  =  2$  (Kallman  \&  Bautista,  2001),  the  ionization
parameter  had been  estimated to  be  log$(\xi) \simeq  4.15$ erg  cm
s$^{-1}$  from the observed  \ion{Fe}{xxv} and  \ion{Fe}{xxvi} absorption
lines.   Since  in  this  case  we did  not  detect  the  \ion{Fe}{xxv}
absorption line we suggested  that the ionization parameter associated
to the \ion{Fe}{xxvi} absorption line was larger than $10^{4.15}$ erg cm
s$^{-1}$.  Furthermore  we were confident  of our result  because Diaz
Trigo  et al.   (2005)  obtained  a similar  value  of the  ionization
parameter (log$(\xi)=4.3 \pm 0.1$  erg cm s$^{-1}$) analysing the same
XMM data of  XB 1254--690 presented above.  Finally  we noted that the
temperature associated  to a  photoionized plasma having  a ionization
parameter  log($\xi$)$> 4.15$  erg cm  s$^{-1}$ should  be $T  \ga 2.4
\times 10^6$ K (see Lee et al., 2002; Kallman \& Bautista, 2001).

At low  energies we detected  
two
weak absorption lines  at 0.835 and 0.849 keV, respectively.
Using   the rest frame  wavelengths reported by Verner et al. (1996) 
we noted that no lines are associated to 0.835 keV, on the other hand 
the   line at  0.849 keV could be associated to   \ion{Fe}{xviii}
and/or  \ion{Fe}{xix} but in this case we should observe more
prominent absorption lines associated to these ions that  were
not observed, for this reason we retained that probably the presence of
these features in the spectra was due to  statistical fluctuations
taking also in account their low statistical significance.

Boirin \&  Parmar (2003) found a  soft excess in the  EPIC pn spectra,
they tried to fit it using  an absorption edge at 1.27 keV as reported
by Iaria et al.  (2001), however the larger spectral resolution of the
EPIC pn indicated  that a Gaussian emission line  centered at 0.93 keV
with a  corresponding equivalent width of  $\sim 30$ eV and  a width $
\sim 175$  eV improved the fit  below 2 keV.  Boirin  \& Parmar (2003)
discussed that  the nature of  the emission feature was  uncertain and
its presence in  the EPIC pn spectra was  almost model independent. We
investigated this feature adding an absorption edge with the threshold
energy fixed to the value reported by Iaria et al.  (2001) and finding
an upper  limit of 0.014 associated  to the optical  depth, this value
was  not compatible  to $0.154  \pm 0.059$  obtained by  Iaria  et al.
(2001).  Moreover we  noted that the soft excess  observed in the EPIC
pn spectra by Boirin \& Parmar (2003) was not present in our residuals
(see Fig, \ref{fig2}).  It is worth to note that a similar soft excess
was also  observed in the  EPIC pn spectrum  of the dipping  source 4U
1323--62 by Boirin  et al.  (2005), in that  case the simultaneous RGS
spectrum did  not show the soft  excess and the  authors discussed the
broad  emission feature  as  a probable  instrumental artifact.   This
conclusion could be valid also for the XMM observation of XB 1254--690
although the lack of a simultaneous  XMM RGS spectrum could not give a
clear confirm of our idea.

We  investigated  the nature  of  the  \ion{Fe}{xxvi}  line width.   A
plausible scenario, already adopted to  describe the line width of the
absorption lines present in the Chandra spectrum of the dipping source
XB 1916--053 (Iaria et al., 2006), is that the line was broadened by
some bulk motion  or supersonic turbulence with a  velocity below 1000
km  s$^{-1}$  as  indicated  by  the  FWHMs of  the  lines  (see  Tab.
\ref{tab1}).  Assuming that the mechanism generating the turbulence or
bulk motion  was due  to the  presence of the  extended corona  we can
achieve  some  informations  about  where the  \ion{Fe}{xxvi} 
 absorption  line  was
produced.  Coronal models tend  to have turbulent velocities which are
locally proportional  to the virial  or rotational velocity  (Woods et
al., 1996).   At $  \sim 8  \times 10^9$ cm  (the coronal  radius, see
Church \& Balucinska-Church, 2004)  the virial velocity should be 1500
km/s, considering  a neutron star  of $1.4 M_{\odot}$.   This velocity
was  similar to the  values obtained  from the  FWHM of  the \ion{Fe}{xxvi}
absorption  line  suggesting that  it  could be
produced between the coronal radius  and the disk edge ($\sim 4 \times
10^{10}$ cm, see Church \& Balucinska-Church, 2004). 
 
\section{Conclusion} 

We  studied the  persistent emission  of XB  1254--690 using  a  53 ks
Chandra observation.  We improved the  position of the source, the new
coordinates  are R.A.=$12^h57^m37^s.153$,  Dec.=$-69^{\circ} 17\arcmin
18\arcsec.98$  (J2000.0) with  an uncertainty  circle of  the absolute
position of 0.6\arcsec.

   We
confirmed  the  presence of  the  \ion{Fe}{xxvi} Ly$_\alpha$  absorption
lines  at   6.962  keV  already  observed  by   XMM-Newton.   Since  a
\ion{Fe}{xxv} absorption  line was not  observed we estimated  that the
ionization parameter log($\xi$) was larger than 4.15 erg cm s$^{-1}$.

The unabsorbed luminosity in the 0.6--10 keV energy range, $1.2 \times
10^{37}$  erg  s$^{-1}$,  similar  to  the values  obtained  from  the
previous  BeppoSAX (Iaria  et al.,  2001) and  XMM (Boirin  \& Parmar,
2003)  observations.  We  estimated  that the  width of the
\ion{Fe}{xxvi} absorption line 
could  be  compatible with  a  broadening  caused  by bulk  motion  or
turbulence  connected  to  the  coronal  activity,  finding  from  the
broadening   that  it  was  produced at  a
distance from the  neutron star between the coronal  radius ($8 \times
10^{9}$ cm) and the disk edge  ($4 \times 10^{10}$ cm).

\begin{acknowledgements}

  This work  was partially  supported by  the Italian
Space  Agency   (ASI)  and  the  Ministero   della  Istruzione,  della
Universit\'a e della Ricerca (MIUR).
\end{acknowledgements}

 {}

\end{document}